# MECHANICALLY MODULATED MICROWAVE CIRCULATOR

*Mustafa Mert Torunbalci[1], Suresh Sridaran[2], Richard C. Ruby[2], and Sunil A. Bhave[1]*
[1]OxideMEMS Lab, Purdue University, West Lafayette, IN, USA
[2]Broadcom Ltd., San Francisco, CA, USA

## ABSTRACT

This work presents a differential FBAR circulator that uses the bending mode to mechanically modulate the FBAR mode without any varactors or switches. The differential FBAR circulator achieves a 61.5 dB isolation (IX) with an insertion loss (IL) of 1.8 dB at 2.68 GHz, demonstrating the first MEMS-only circulator. The isolation bandwidth at -25 dB is 4.7 MHz and power handling of the circulator is limited by the FBARs to +34 dBm.

## INTRODUCTION

Circulators are three-port, non-reciprocal devices that transmit RF signal from one port to the next one, only in one direction. These devices are key components for achieving full duplex wireless systems by enabling simultaneous transmission and reception. A common way to realize a circulator is to use an external magnetic field in a ferromagnetic material. However, the ferromagnetic films are not CMOS process compatible and the technique does not scale favorably with volume reduction. Magnetic-free circulators have been demonstrated by spatio-temporal modulation of a resonant cavity using varactors or switches [1-5]. Further improvements in the insertion loss (IL), isolation bandwidth (BW), and linearity (no side bands) have been reported by using differential modulation of LC tanks or FBARs [3,6,7]. However, these techniques are still limited in size, linearity and RF power handling by the external varactors or switches.

In this paper, we present a MEMS-only circulator that uses its bending mode to mechanically modulate the FBAR mode. The use of bending mode for modulation eliminates the need for external tuning elements such as varactors or switches, significantly simplifying the modulation network, improving the linearity and RF power handling, and reducing the size.

## ARCHITECTURE

Figure 1 presents the block diagram of the proposed mechanically modulated FBAR circulator. FBARs are modeled with an equivalent BVD circuit having two branches for bending and FBAR modes. The FBAR's mechanical compliance (modeled by $C_{x1}$) is a function of the motional charge ($q_m$) of the bending mode. The bending mode is used to mechanically modulate the fundamental FBAR mode with appropriate phase difference. Three such FBARs make up a single-ended FBAR circulator chip (Figure 1.b). Two chips are connected in parallel and driven in differential spatio-temporal modulated signal to form the pseudo linear-time-invariant (LTI) mechanical circulator. Figure 2 shows the pictures of the FBAR circulator. The circulator uses two FBAR chips implemented in a differential topology. Each FBAR chip includes three identical FBARs with a series frequency of 2.65 GHz connected to a common node. The top AlN layer on the FBAR stack is used to break the symmetry to provide a bending mode.

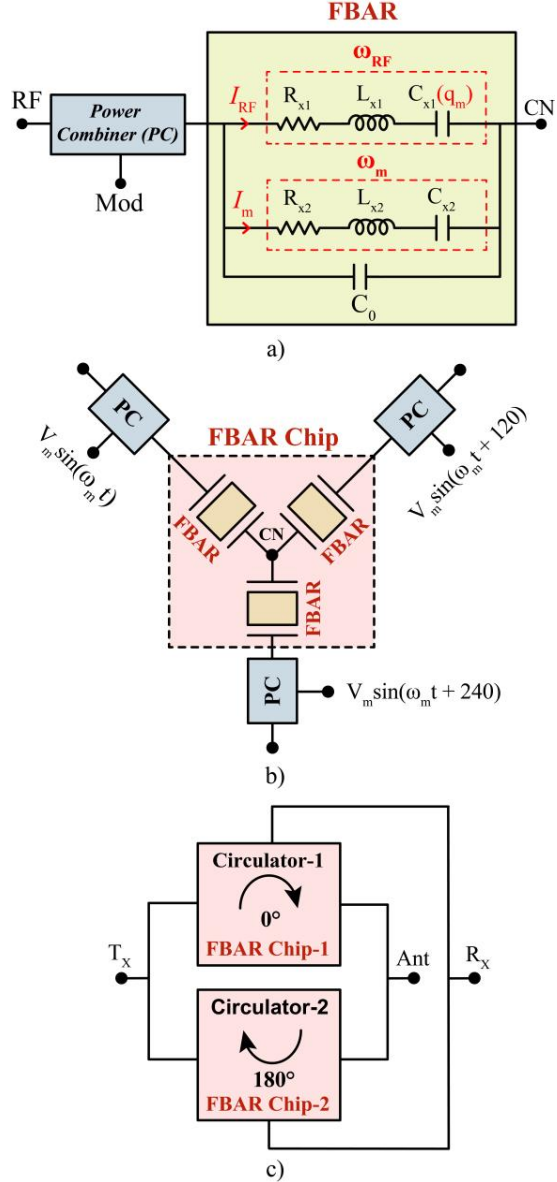

*Figure 1: Block diagram of mechanically modulated circulator: a) Equivalent circuit model of a single FBAR branch, showing two BVD branches corresponding to the FBAR and bending mode. b) single-ended circulator with a single FBAR chip. c) differential circulator with two FBAR chips.*

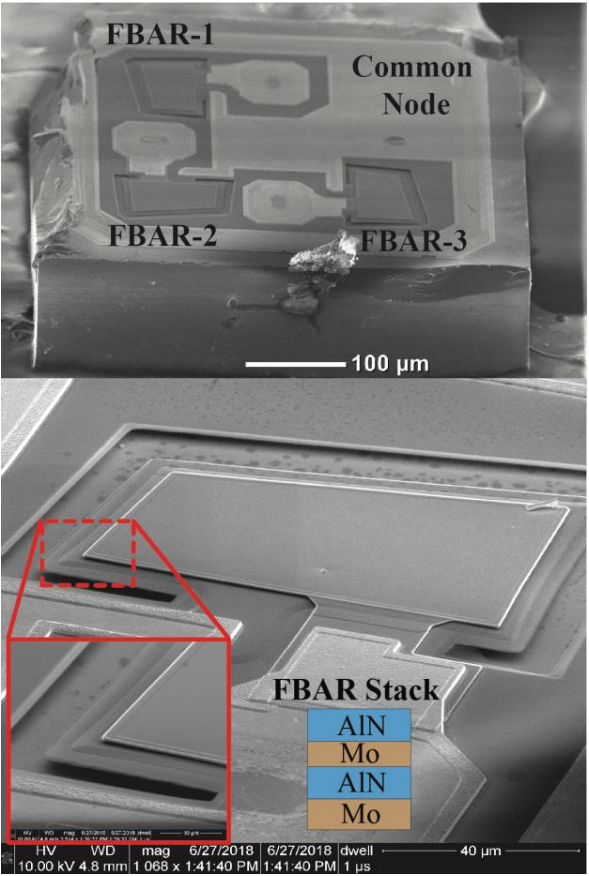

*Figure 2: SEMs of the film bulk acoustic resonator chip. Each chip is 0.05 mm³. Two chips with total 6 FBARs are used to implement differential circulator on a PCB.*

Bending stress generated on the FBAR shifts both series and parallel frequency [8] and provide an isolation when modulated with proper phase difference in the time domain.

## FBAR

The broadband response of a single FBAR is measured using an Agilent 3-port network analyzer. Figure 3 shows measured broadband response, showing the fundamental and bending modes. The BVD parameters are extracted using this sensor. FBARs have a Q of 700 and piezoelectric coupling coefficient of 9%.

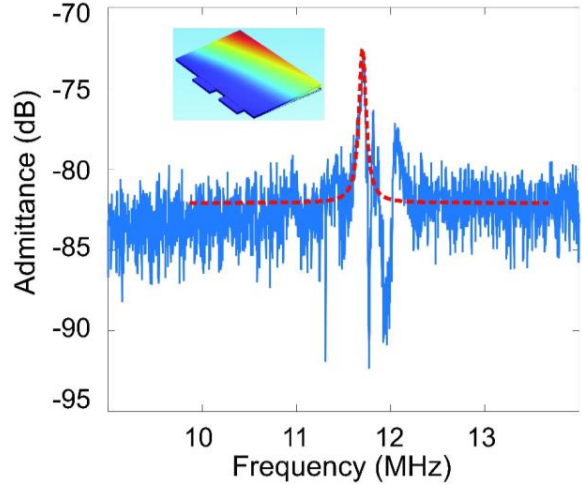

*Figure 4: Zoom-in scan and measurement of the low frequency bending modes of an FBAR. Since the FBAR is almost symmetric in thickness, AlN only couples weakly to the bending modes. The resonators operate in air where the mechanical bending mode is severely damped.*

We perform very low frequency modal spectroscopy by taking very careful low noise measurement and identify the electromechanical admittance response of the FBAR's bending mode (Figure 4). A Lorentzian function is fitted to the response of bending mode and $Q_{bending}$ is calculated to be 100. While a high $Q_{bending}$ would significantly reduce the modulator drive voltage, a Q of 100 is a good compromise because of small difference in the bending mode frequencies of the 3 FBARs.

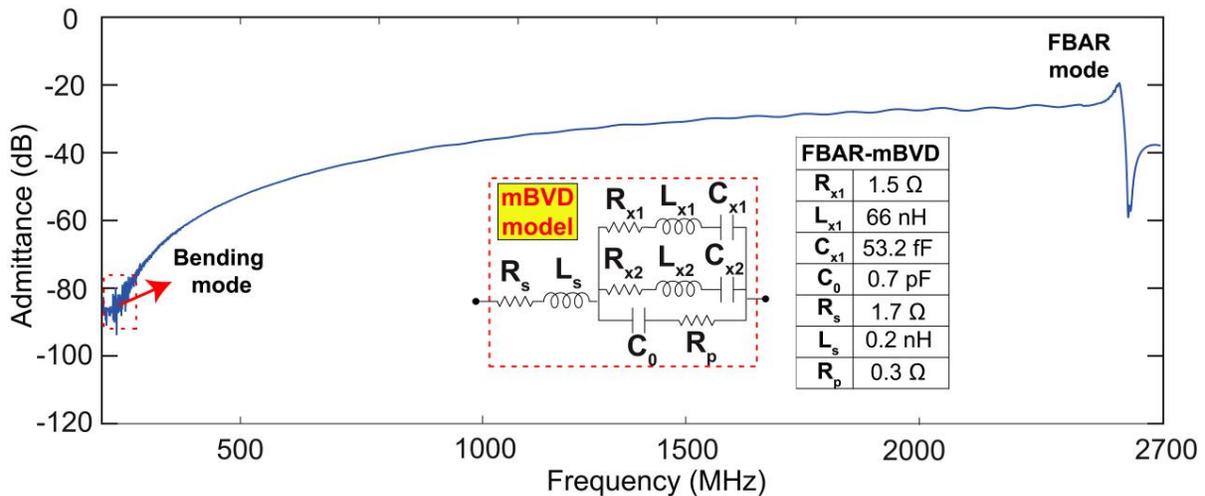

*Figure 3: Measured broad-band response of a single FBAR showing the fundamental bulk acoustic mode and extremely small signals corresponding to the bending mode. The Butterworth-Van-Dyke (BVD) parameters are extracted using this resonator.*

## CIRCULATOR

The circulator uses 2 FBAR chips each having three FBARs, implanted in differential architecture (Figure 5). Unlike the multi-point anchors of standard apodized FBARs, the designs for the circulator are cantilevered single anchor FBARs at 2.65 GHz. Driving the bending mode at 11.6 MHz, which we refer to as the modulation mode, generates dynamic stress in the resonator which changes the stiffness of the FBAR mode. This results in modulation of the FBAR's series and parallel resonance [8].

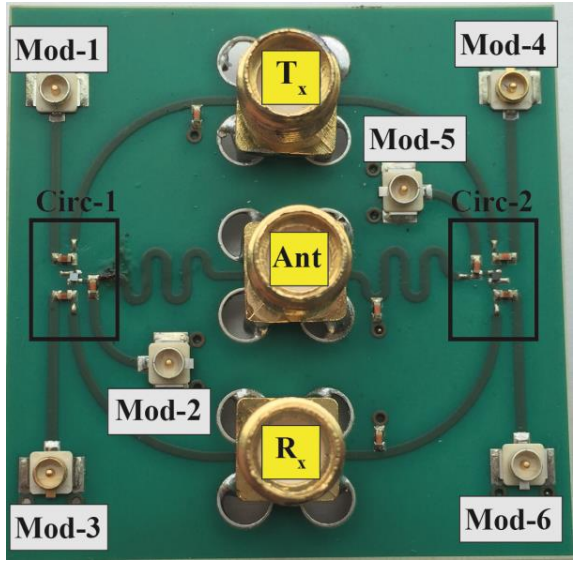

*Figure 5: 99.99% of the circulator board volume is occupied by the SMA connectors for providing 3 RF and 6 modulation signals and RF power splitter T-lines. Each FBAR chip is 0.5mm x 0.5mm x 0.2mm, resulting in a total chip volume of 0.1 mm³.*

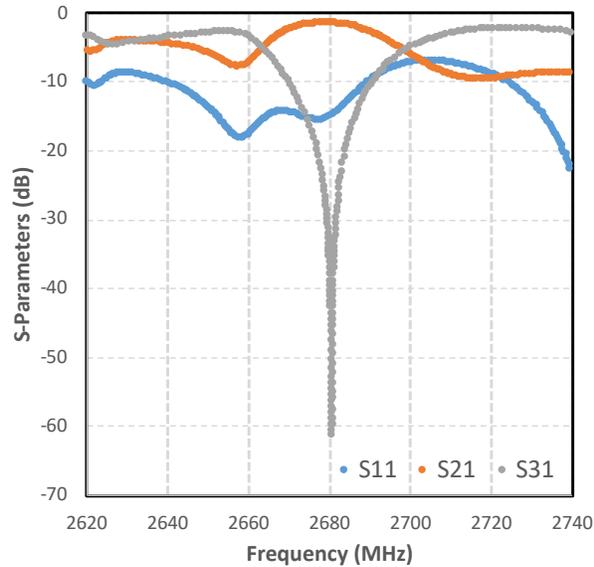

*Figure 6: Measured S-parameter response of the all-mechanical differentially modulated circulator. The circulator achieves a 61.5 dB isolation (IX) with an insertion loss (IL) of 1.8 dB at 2.68 GHz.*

Circulation is achieved by modulating all 6 FBARs with the correct phase signals (Figure 6). When the modulation signals are off, power is split equation from port 1 to ports 2 and 3. However when modulated at the bending mode frequency, the modulation amplitude is amplified by $Q_{bending} = 100$, providing sufficient stress to shift the frequency of the FBAR mode. The mechanical bending reaches maximum displacement twice in one oscillation cycle. Therefore, the FBAR is modulated at twice the bending mode frequency (Table 1).

*Table 1: Summary of results and comparison with previous works.*

| Metric | [2] | This Work |
|---|---|---|
| *Technology* | FBAR | FBAR |
| *Frequency (MHz)* | 2500 | 2680 |
| *Modulation* | Varactor | Mechanical |
| $f_{mod}$ *(MHz)* | 3 | 11.6<br>23.2 to FBARs |
| *IX (dB)* | 76 | 61.5 |
| *IL (dB)* | 11 | 1.8 |
| *BW@25dB IX (MHz)* | 1 | 4.7 |

## CONCLUSIONS

This work demonstrates a MEMS-only solution for the implementation of a microwave circulator by mechanically modulating FBAR mode with proper phase difference using its bending mode. The differential FBAR circulator achieves an isolation of -61.5 dB at 2.68 GHz with an insertion loss and isolation bandwidth of 1.8 dB and 4.7 MHz, respectively. The non-linearity is no longer limited by switches and varactors, but by the FBAR itself. The mechanically modulated microwave circulator has eliminated the need for off-chip, non-linear, active components, making the FBAR circulator an attractive candidate for low-temperature, magnet-free isolators. The current design needs 6 phases of the modulation clock. While these can be provided by a CMOS delay-line oscillator chip, future designs will explore delivering the different phases of the modulation clock using an ultrasonic traveling wave [9].

## ACKNOWLEDGEMENTS

The authors would like to acknowledge support from ONR Quantum Information Systems program and DARPA MTO Seedling. We thank Michael Quinn of Majelac Inc for PCB assembly. We also want to thank Trevor J. Odelberg and Dr. Mohammed Abu Khater for their invaluable input on PCB design, assembly and corrections.

No.5, pp. 395-397, 2018.

**CONTACT**

M. Mert Torunbalci, mtorunba@purdue.edu
S. Bhave, bhave@purdue.edu